\documentclass[a4paper,10pt]{article}
\begin{document}
\title{Matrix Product Steady States as Superposition of Product Shock Measures in 1D Driven Systems}%
\author{F H Jafarpour$^{1}$\footnote{Corresponding author's e-mail:farhad@ipm.ir} \, and S R Masharian$^{2}$ \\ \\%
{\small $^1$Bu-Ali Sina University, Physics Department, Hamadan, Iran} \\%
{\small $^2$Institute for Advanced Studies in Basic Sciences, Zanjan, Iran}}%
\maketitle %
\begin{abstract}
It is known that exact traveling wave solutions exist for families
of $(n+1)$-states stochastic one-dimensional non-equilibrium lattice
models with open boundaries provided that some constraints on the
reaction rates are fulfilled. These solutions describe the diffusive
motion of a product shock or a domain wall with the dynamics of a
simple biased random walker. The steady state of these systems can
be written in terms of linear superposition of such shocks or domain
walls. These steady states can also be expressed in a matrix product
form. We show that in this case the associated quadratic algebra of
the system has always a two-dimensional representation with a
generic structure. A couple of examples for $n=1$ and $n=2$ cases
are presented.
\end{abstract}
\maketitle %
On the macroscopic level nonlinear hydrodynamic equations, such as
the Burgers equation or the Fisher equation, might exhibit shocks in
some cases \cite{bu,fi}. On the microscopic level some stochastic
non-equilibrium one-dimensional lattice models with open boundaries
might also develop shocks provided that some constrained on the
reaction rates are fulfilled \cite{sp,re,kl}.\\
In \cite{kjs} exact traveling wave solutions have been obtained for
three families of two-states driven-diffusive models in
one-dimension with nearest-neighbors interactions. These three
models are the Partially Asymmetric Simple Exclusion Process (PASEP)
\cite{li,sch,dehp}, the Branching-Coalescing Random Walk (BCRW) and
the Asymmetric Kawasaki-Glauber Process (AKGP). By taking a
Bernoulli shock distribution as initial distribution, the shock
distribution in these models evolves in time into a linear
combination of similar distributions with different shock positions.
The time evolution of the shock position in these models is similar
to the dynamics of a random walker moving on a one-dimensional
lattice with reflecting boundaries \cite{kjs}. This property of the
traveling wave solution is not limited to the two-states systems
with nearest-neighbor interactions.\\
Shocks in three-states lattice models with nearest-neighbor
interactions \cite{jm,rs,ja1,ts1,ts2} and also two-states lattice
models with next nearest-neighbors interactions \cite{ps} have also
been studied recently. As long as the time evolution of the product
shock measure is equivalent to that of a random walker on a discrete
lattice with homogeneous hopping rates in the bulk and special
reflection rates at the boundary, the steady state of the model can
be written as a linear superposition of these product shock
measures.\\
The steady states of these systems, on the other hand,
can be obtained using the Matrix Product Formalism (MPF) first
introduced in \cite{dehp} and then developed in \cite{ks} (for a
recent review see \cite{be}). According to the MPF the steady state
weight of a given configuration of a two-states stochastic
non-equilibrium one-dimensional lattice models with open boundaries
can be written as expectation value of product of non-commuting
operators associated with different types of particles and vacancies
in that configuration. These operators should satisfy an associative
algebra. By using a matrix representation of this algebra or using
the commutation relations of the operators one can in principle
calculate the unnormalized steady state weights. Whether or not this
algebra has a finite- or infinite-dimensional representation in not
known in general; however, it is known that some quadratic algebras
have finite-dimensional representations on special manifolds in
their parameters space. For instance it has been shown that the
quadratic algebra of PASEP has finite-dimensional representations
given that one is restricted to some special values of the reaction
rates through some constraints \cite{er,ms}. Interestingly, it has
been shown that the constraints for the existence of a
two-dimensional representation are exactly those necessary for a
single product shock measure as an initial configuration to have a
simple random walk dynamics on the lattice \cite{bs}. It has also
been shown that if one takes a product measure of $N-1$ consecutive
product shocks, the shock positions have simple random walk dynamics
provided that the constraints for the existence of an
$N$-dimensional representation for the quadratic algebra of the
PASEP are fulfilled. To see this one should adopt a slightly
different definition of the shock measure. According to this new
definition the shock is a product measure with density $1$ at the
shock positions and intermediate densities between these sites
\cite{bs}. Hence the $N$-dimensional representations of the
stationary algebra describe the stationary linear combination of
such shock measures. The same is true for both the BCRW and the AKGP
\cite{kjs}. In fact it is known that the quadratic algebras
associated with the BCRW and that of the AKGP have two-dimensional
representations under some conditions on the reaction rates.
Surprisingly, it has been shown that a single product measure in
these systems will have a random walk dynamics under exactly the
same constraints \cite{ja2}. In \cite{rs} a three-states model with
open boundary has been studied and it has been shown that the
traveling shock solutions exist under some constraints on the
reaction rates. Later a generalization of this model was introduced
and studied in \cite{ts1}. The authors in \cite{jm} have shown that
the quadratic algebra of this generalized model can be mapped into
the quadratic algebra of the PASEP provided that the constraints
under which the traveling shock solutions exist, are fulfilled. This
means that the quadratic algebra of this generalized model has two
dimensional
representation under the same constraints.\\
Our major motivation in this paper is to investigate how the steady
states of one-dimensional driven-diffusive systems with open boundaries
when written in terms of linear superposition of Bernoulli shocks are related
to the matrix product steady states. In this direction, we will consider those
$(n+1)$-states stochastic non-equilibrium one-dimensional lattice models with
open boundaries and nearest-neighbors interactions in which the time evolution
equation of a single product shock measure is similar to the evolution equation
for a biased single-particle random walk moving on a finite lattice with reflecting
boundaries. Since the dynamics of the shock position, as a single-particle excitation,
is simply a random walk it is easy to find the steady state probability distribution
function of these systems in terms of linear superposition of single shocks. We will
then show that the matrix product steady state of these systems can always be obtained
using two-dimensional representations of their quadratic algebras. These matrix representations
have always a generic form regardless of the details of the microscopic reactions. In order to
confirm our assertion we will give a couple of example for both $n=1$ and $n=2$ cases. \\
In what follows we will first review very quickly the dynamics of a biased single-particle
random walk on a discrete lattice of finite length and show how the steady states probability
distribution of this system can be obtained. The time evolution of the probability distribution
function of any configuration of a Markovian interacting particle system $\vert P(t)\rangle$
is governed by a master equation which can be written as a Schr\"odinger like equation in imaginary time
\begin{equation}
\label{ME}
\frac{d}{dt}\vert P(t) \rangle =H \vert P(t) \rangle
\end{equation}
in which $H$ is a stochastic Hamiltonian \cite{sch}. The matrix
elements of the Hamiltonian are the transition rates between
different configurations. For the one-dimensional systems defined on
a lattice of length $L$ with nearest neighbors interactions the
Hamiltonian $H$ has the following general form
\begin{equation}
\label{Hamiltonian} H=\sum_{k=1}^{L-1}h_{k,k+1}+h_1+h_L.
\end{equation}
in which
\begin{eqnarray}
\label{Hamiltoniandet} h_{k,k+1}&=&{\cal I}^{\otimes (k-1)}\otimes h
\otimes {\cal I}^{\otimes (L-k-1)} \\ h_1 &=& h^{(l)} \otimes {\cal
I}^{\otimes (L-1)}
\\ h_L &=& {\cal I}^{\otimes (L-1)}\otimes h^{(r)}
\end{eqnarray}
For an ($n+1$)-states one-dimensional driven diffusive system, in
which these states are associated with $n$ different types of
particles and empty sites, ${\cal I}$ is an $(n+1) \times (n+1)$
identity matrix and $h$ is an $(n+1)^2 \times (n+1)^2$ matrix for the
bulk interactions. The two matrices $h^{(r)}$ and $h^{(l)}$ are both
$(n+1) \times (n+1)$ square matrices and determine the interactions at the
boundaries.\\
Let us assume that at $t=0$ the probability distribution function of
the an $(n+1)$-states one-dimensional driven diffusive system is
given by a product shock measure defined on a lattice of length $L$
as follows
\begin{equation}
\label{PSM}
\vert k \rangle= \left( \begin{array}{c} \rho_1^{(0)} \\
\rho_1^{(1)} \\ \vdots \\
\rho_1^{(n)} \end{array} \right)^{\otimes k} \otimes \left( \begin{array}{c} \rho_2^{(0)} \\
\rho_2^{(1)} \\ \vdots \\
\rho_2^{(n)} \end{array} \right)^{\otimes L-k} \; \;,\; \; 0\leq k
\leq L
\end{equation}
in which the density of particles of type $i$ ($i=1,\cdots,n$) on
the left (right) hand side of the shock position is $\rho_{1}^{(i)}$
($\rho_{2}^{(i)}$). The density of the empty sites on the left and
right hand sides of the shock position are given by
$\rho_{1}^{(0)}=1-\sum_{i=1}^{n}\rho_{1}^{(i)}$ and
$\rho_{2}^{(0)}=1-\sum_{i=1}^{n}\rho_{2}^{(i)}$ respectively. As can
be seen every species of particles has a shock profile and the
centers of shocks lie on each other. We assume that under some
constraints on the reaction rates of the system the dynamics of this
state $\vert k \rangle$, which is given by (\ref{ME}), is similar to
the dynamics of a biased single-particle random walker on a discrete
lattice with reflecting boundaries. In this case, regardless of
particle type, the shock position hops to the left and right with
the rates $\delta_{l}$ and $\delta_{r}$ respectively. The shock
position also reflects from the boundaries. It reflects from the
left boundary with the rate $\bar \delta_{r}$ and from the right
boundary with the rate $\bar \delta_{l}$. The evolution of the state
$\vert k \rangle$ is then given by
\begin{eqnarray}
\label{RWE1} H \vert k \rangle &=& \delta_{l} \vert k-1 \rangle +
\delta_{r} \vert k+1
\rangle - (\delta_{l}+\delta_{r}) \vert k \rangle, 0 < k < L \\
\label{RWE2}
H \vert 0 \rangle &=& \bar \delta_{r} \vert 1\rangle -\bar \delta_{r}\vert 0 \rangle,\\
\label{RWE3} H \vert L \rangle &=& \bar \delta_{l} \vert L-1\rangle
-\bar \delta_{l}\vert L \rangle.
\end{eqnarray}
Now one can construct a linear superposition of these single shocks
to make the steady state of the system $\vert P^*\rangle$ as follows
\begin{equation}
\label{SSS} \vert P^* \rangle =\sum_{k=0}^{L}c_{k}\vert k \rangle
\end{equation}
so that it satisfies
\begin{equation}
\label{SSC} H \vert P^* \rangle =0.
\end{equation}
Using (\ref{RWE1})-(\ref{SSC}) one can easily find the coefficients $c_{k}$'s as
\begin{eqnarray}
\label{COEF1}
c_{k}&=&\frac{({\delta_{r}}/{\delta_{l}})^k}{[({\delta_{r}}/{\bar
\delta_{r}})+
\frac{({\delta_{r}}/{\delta_{l}})}{1-({\delta_{r}}/{\delta_{l}})}]
+[({\delta_{l}}/{\bar \delta_{l}})-
\frac{1}{1-({\delta_{r}}/{\delta_{l}})}]({\delta_{r}}/{\delta_{l}})^L} , 0 < k < L\\
\label{COEF2}
c_{0}&=&\frac{({\delta_{r}}/{\bar\delta_{r}})}{[({\delta_{r}}/{\bar
\delta_{r}})+
\frac{({\delta_{r}}/{\delta_{l}})}{1-({\delta_{r}}/{\delta_{l}})}]
+[({\delta_{l}}/{\bar \delta_{l}})-
\frac{1}{1-({\delta_{r}}/{\delta_{l}})}]({\delta_{r}}/{\delta_{l}})^L},\\
\label{COEF3}
c_{L}&=&\frac{({\delta_{l}}/{\bar\delta_{l}})({\delta_{r}}/{\delta_{l}})^L}
{[({\delta_{r}}/{\bar\delta_{r}})+
\frac{({\delta_{r}}/{\delta_{l}})}{1-({\delta_{r}}/{\delta_{l}})}]
+[({\delta_{l}}/{\bar \delta_{l}})-
\frac{1}{1-({\delta_{r}}/{\delta_{l}})}]({\delta_{r}}/{\delta_{l}})^L}.
\end{eqnarray}
The equation (\ref{SSS}) gives the steady state probability
distribution function of the system in terms of superposition of
product shock measures. Because of the uniqueness of the steady
state of our system, one should be able to find the same
distribution function using other methods such as the MPF. In what
follows we will briefly review the basic concepts of the MPF for an
($n+1$)-states system. \\
According to the MPF the stationary probability distribution $\vert
P^*\rangle$ for a system of length $L$ is assumed to be of the form
\begin{equation}
\vert P^* \rangle =\frac{1}{Z} \langle W \vert \left( \begin{array}{c} D_{0} \\
D_{1} \\ \vdots \\
D_{n} \end{array} \right)^{\otimes L} \vert V \rangle
\end{equation}
in which $Z$ is a normalization factor and can be obtained easily
using the normalization condition to be
\begin{equation}
\label{NOR}
Z=\langle W \vert (\sum_{i=0}^{n}D_{i})^L\vert V \rangle
\end{equation}
The operator $D_{0}$ stands for the presence of a vacancy and
the operator $D_{i}$ ($i=1,\cdots,n$) stands for the presence of a
particle of type $i$ at each lattice site. These operators besides the
vectors $\langle W \vert$ and $\vert V \rangle$ should satisfy the
following algebra
\begin{equation}
\label{Twoalg}
\begin{array}{l}
h \left[ \left( \begin{array}{c} D_{0} \\
D_{1} \\ \vdots \\
D_{n} \end{array} \right) \otimes
\left( \begin{array}{c} D_{0} \\
D_{1} \\ \vdots \\
D_{n} \end{array} \right) \right]=
\left( \begin{array}{c} \bar{D_{0}} \\
\bar{D_{1}} \\ \vdots \\
\bar{D_{n}} \end{array} \right) \otimes
\left( \begin{array}{c} D_{0} \\
D_{1} \\ \vdots \\
D_{n} \end{array} \right) -
\left( \begin{array}{c} D_{0} \\
D_{1} \\ \vdots \\
D_{n} \end{array} \right) \otimes
\left( \begin{array}{c} \bar{D_{0}} \\
\bar{D_{1}} \\ \vdots \\
\bar{D_{n}} \end{array} \right), \\
\langle W| \; h^{(l)} \left( \begin{array}{c} D_{0} \\
D_{1} \\ \vdots \\
D_{n} \end{array} \right) =
-\langle W| \left( \begin{array}{c} \bar{D_{0}} \\
\bar{D_{1}} \\ \vdots \\
\bar{D_{n}} \end{array} \right),
h^{(r)} \left( \begin{array}{c} D_{0} \\
D_{1} \\ \vdots \\
D_{n} \end{array} \right) |V \rangle =
\left( \begin{array}{c} \bar{D_{0}} \\
\bar{D_{1}} \\ \vdots \\
\bar{D_{n}} \end{array} \right) |V \rangle \nonumber
\end{array}
\end{equation}
in which $\bar{D_{i}}$'s are auxiliary operators \cite{ks,be}. These relations
define a quadratic algebra. Using the algebra or one of its matrix
representations, one can calculate the probability of any
configuration in the steady state. It is also possible to calculate
the mean values of physical quantities such as the current of
particles of different species or the correlation functions in the
steady state.\\
As we mentioned above, by requiring some constraints on the reaction
rates of the system the dynamics of a single shock can simply be
given by a single random walk dynamics. From there one can
construct the steady state of the system. We should note that the
results obtained from the random walk picture should be equal to
those obtained from the MPF and in fact we should have
\begin{equation}
\label{EQU}
\frac{1}{Z} \langle W \vert \left( \begin{array}{c} D_{0} \\
D_{1} \\ \vdots \\
D_{n} \end{array} \right)^{\otimes L} \vert V \rangle=
\sum_{k=0}^{L}c_{k}\vert k \rangle.
\end{equation}
Let us assume that we have a matrix representation for the operators
$D_{i}$'s and the vectors $\vert V \rangle$ and $\langle W \vert$.
Finding the bra $\langle k' \vert$ which is orthonormal to $\vert k
\rangle$ i.e. $\langle k' \vert k \rangle =\delta_{k'k}$ helps us
calculate the coefficients $c_k$. By defining
\begin{equation}
\vert k' \rangle =\left( \begin{array}{c} x_{0} \\
x_{1} \\ \vdots \\
x_{n} \end{array} \right)^{\otimes k'} \otimes\left( \begin{array}{c} y_{0} \\
y_{1} \\ \vdots \\
y_{n} \end{array} \right)^{\otimes L-k'}
\end{equation}
and requiring the orthonormality condition one finds that only four
of $2n+2$ parameters $\{x_{0},\cdots,x_{n},y_{0},\cdots,y_{n}\}$ are
independent. One can simply find a solution by taking
$x_0=\frac{\rho_2}{\rho_2-\rho_1}$,
$x_{1}=\cdots=x_{n}=\frac{-(1-\rho_2)}{\rho_2-\rho_1}$,
$y_0=\frac{-\rho_1}{\rho_2-\rho_1}$ and
$y_{1}=\cdots=y_{n}=\frac{(1-\rho_1)}{\rho_2-\rho_1}$ in which
$\rho_{1}=\sum_{i=1}^{n}\rho_{1}^{(i)}$ and
$\rho_2=\sum_{i=1}^{n}\rho_{2}^{(i)}$ are the total density of
particles on the left and the right hand sides of the shock position
respectively. By multiplying $\langle k' \vert$ in (\ref{EQU}) from
the left one finds
\begin{equation}
\label{COEF4} c_{k}=\frac{1}{Z}\langle W \vert
G_{1}^{k}G_{2}^{L-k}\vert V \rangle
\end{equation}
in which
\begin{equation}
G_{1}=\sum_{i=0}^{n}x_{i}D_{i}\;\;\;\mbox{and}\;\;\;\
G_{2}=\sum_{i=0}^{n}y_{i}D_{i}.
\end{equation}
We have found that the following two-dimensional representation
\begin{equation}
\label{REP}
\begin{array}{l}
D_{0}=\left( \begin{array}{cc}
(1-\sum_{i=1}^{n}\rho_{2}^{(i)}) & 0\\
d_0 & \frac{\delta_{r}}{\delta_{l}}(1-\sum_{i=1}^{n}\rho_{1}^{(i)})\\
\end{array} \right),\\
D_{i}=\left( \begin{array}{cc}
\rho_{2}^{(i)} & 0\\
d_{i} & \frac{\delta_{r}}{\delta_{l}}\rho_{1}^{(i)}\\
\end{array} \right) \;\;\; i=1,\cdots,n \;\;,\\
\vert V \rangle =\left( \begin{array}{c}
v_1\\
v_2\\
\end{array} \right) \;,\; \langle W \vert=\left( \begin{array}{cc}
w_1&w_2\\
\end{array} \right)
\end{array}
\end{equation}
in which $d_{0}=-\sum_{i=1}^{n}d_{i}$, uniquely generates the same
coefficients $c_k$ for $k=0,\cdots,L$ as in
(\ref{COEF1})-(\ref{COEF3}) provided that
\begin{eqnarray}
d_{0}\frac{w_2}{w_1}=(\rho_2-\rho_1)\frac{(\frac{\bar
\delta_r}{\delta_l})}{1-(\frac{\delta_r}{\delta_l})+(\frac{\bar\delta_r}{\delta_l})},\\
d_{0}\frac{v_1}{v_2}=(\rho_1-\rho_2)\frac{(\frac{\bar
\delta_l}{\delta_l})}{1-(\frac{\delta_l}{\delta_r})+(\frac{\bar\delta_l}{\delta_r})}.
\end{eqnarray}
These two equations tell us how the reflection rates $\bar
\delta_{r}$ and $\bar \delta_{l}$ are related to the components of
the two vectors $\vert V \rangle$ and $\langle W \vert$. One should
note that in (\ref{REP}), $d_{i}$'s ($i=1,\cdots,n$) are free
parameters. Since the matrix $\sum_{i=0}^{n}D_{i}$ is diagonal in
this representation, the normalization factor $Z$ in (\ref{NOR}) can
easily be calculated as
\begin{equation}
Z=w_1 v_1 + (\frac{\delta_{r}}{\delta_{l}})^L w_2 v_2.
\end{equation}
The matrix representations for the two matrices $G_{1}$ and $G_{2}$
are now given by
\begin{equation}
G_{1}=\left( \begin{array}{cc}
0 & 0\\
\frac{d_0}{\rho_{2}-\rho_{1}}&\frac{\delta_{r}}{\delta_{l}}\\
\end{array} \right),\\
G_{2}=\left( \begin{array}{cc}
1 & 0\\
\frac{-d_0}{\rho_{2}-\rho_{1}}& 0\\
\end{array} \right)
\end{equation}
with these properties
\begin{equation}
G_{1}^{k}=(\frac{\delta_{r}}{\delta{l}})^{k-1}G_{1}\;\;,\;\;G_{2}^{k}=G_{2}.
\end{equation}
In summary whenever the quadratic algebra of an ($n+1$)-states
system has a two-dimensional representation of the form (\ref{REP})
one can calculate the coefficients $c_{k}$ using (\ref{COEF4}) and
conclude that the steady state of the system is a linear
superposition of single product shock measures with random walk
dynamics. In order to confirm our assertion we will give a couple of
examples. \\
Let us first study the shocks in two-states systems which can be
interpreted as particles (denoted by $1$) and vacancies (denoted by
$0$) at each lattice site. The stochastic dynamics are defined in
terms of transition rates. In this case the process is fully defined
by the 12 rates $w_{ij}$ ($i\neq j$) where $i,j=1,\cdots,4$. In the basis
($00,01,10,11$) the reactions in the bulk of the system are as follows
\begin{eqnarray}
01 \rightleftharpoons 10 \quad & w_{32}, \; w_{23} \nonumber \\
11 \to 10,01 \quad & w_{34}, \; w_{24} \nonumber \\
10,01 \to 11 \quad & w_{43}, \; w_{42} \nonumber \\
10,01 \to 00 \quad & w_{13}, \; w_{12} \nonumber \\
00 \to 10,01 \quad & w_{31}, \; w_{21} \nonumber \\
11 \rightleftharpoons 00  \quad & w_{14}, \; w_{41}. \nonumber
\end{eqnarray}
For injection and extraction of particles at the boundaries we
introduce the four rates $\alpha$, $\beta$, $\gamma$ and $\delta$.
The reactions at the left and the right boundaries are
\begin{eqnarray}
0\rightleftharpoons 1 \quad & \alpha, \; \gamma \quad \mbox{and} \nonumber \\
0\rightleftharpoons 1 \quad & \delta, \; \beta   \nonumber
\end{eqnarray}
respectively.
As the first example we consider the PASEP. It is well known that
the steady state of the PASEP can be written in terms of
superposition of shocks with random walk dynamics provided that some
constraints on the reaction and the boundaries rates are fulfilled. In the bulk of
the lattice the particles diffuse to the left and right with the
rates $w_{32}=x$ and $w_{23}=1$ respectively.
For our continence we consider the injection and extraction rates for
the left boundary as $(1-x)\alpha$ and $(1-x)\gamma$ and also for
right boundary as $(1-x)\delta$ and $(1-x)\beta$. The associated
quadratic algebra of this model can now be written as
\begin{equation}
\label{PASEPalg}
\begin{array}{l}
D_{1}D_{0}-x  D_{0}D_{1}=\xi(1-x)(D_{0}+D_{1})\\
\langle W \vert (\alpha D_{0} - \gamma D_{1}- \xi)=0\\
(\beta D_{1}- \delta D_{0}-\xi)\vert V \rangle=0.
\end{array}
\end{equation}
It is assumed that $\bar {D_{0}}=\xi(1-x) \mathcal{I}$ and $\bar
{D_{1}}=-\xi(1-x) \mathcal{I}$ in which $\mathcal{I}$ is a $2\times
2$ identity matrix. One can easily check the following
two-dimensional representation satisfies (\ref{PASEPalg})
\begin{equation}
D_{0}=\left( \begin{array}{cc}
(1-\rho_{2}) & 0\\
d_{0}&\frac{\delta_{r}}{\delta_{l}}(1-\rho_{1})\\
\end{array} \right),\\
D_{1}=\left( \begin{array}{cc}
\rho_{2} & 0\\
-d_0& \frac{\delta_{r}}{\delta_{l}}\rho_{1}\\
\end{array} \right)
\end{equation}
provided that
\begin{equation}
x=\frac{\rho_{1}(1-\rho_{2})}{\rho_{2}(1-\rho_{1})}\;\; ,\;\;
\frac{\delta_{r}}{\delta_{l}}=\frac{\rho_{2}(1-\rho_{2})}{\rho_{1}(1-\rho_{1})}
\end{equation}
and $\xi=\rho_{2}(1-\rho_{2})$. The boundary rates together with
$\rho_{1}$, $\rho_{2}$ and $x$ should also satisfy the following
conditions
\begin{equation}
\label{PASEPcond}
\begin{array}{l}
\alpha(1-\rho_1)-\gamma\rho_1=\rho_1(1-\rho_1)(1-x) \\
\beta \rho_2-\delta(1-\rho_2)=\rho_2(1-\rho_2)(1-x).
\end{array}
\end{equation}
In this case none of the elements $v_{1}$, $v_{2}$, $w_{1}$ and
$w_{2}$ in the vectors $\vert V \rangle$ and $\langle W \vert$ is
zero. Finite-dimensional representations of the PASEP have widely
been studied in \cite{er,ms}.
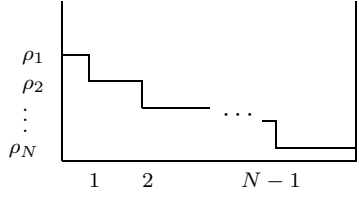
\begin{figure}
\begin{center}
\begin{picture}(120,70)
\put(10,10){\line(0,1){60}}
\put(10,10){\line(1,0){110}}
\put(120,10){\line(0,1){60}}
\put(10,50){\line(1,0){10}}
\put(20,50){\line(0,-1){10}}
\put(20,40){\line(1,0){20}}
\put(40,40){\line(0,-1){10}}
\put(40,30){\line(1,0){25}}
\put(85,25){\line(1,0){5}}
\put(70,25){$\cdots$}
\put(90,15){\line(0,1){10}}
\put(90,15){\line(1,0){30}}
\put(20,0){\footnotesize $1$}
\put(40,0){\footnotesize $2$}
\put(77,0){\footnotesize $N-1$}
\put(-10,13){\footnotesize $\rho_{N}$}
\put(-5,21){\footnotesize $\vdots$}
\put(-5,37){\footnotesize $\rho_{2}$}
\put(-5,48){\footnotesize $\rho_{1}$}
\end{picture}
\caption[fig]{Sketch of a multiple shock structure in the PASEP with open boundaries.
The shock positions are at the sites $1$, $2$, $\cdots$, and $N-1$.}
\label{fig}
\end{center}
\end{figure}
It is known that the PASEP has also an $N$-dimensional representation provided that
\begin{eqnarray}
x^{1-N}=\kappa_{+}(\beta,\delta)\kappa_{+}(\alpha,\gamma)
\end{eqnarray}
in which
\begin{equation}
\label{kappa}
\kappa_{+}(u,v)=\frac{-u+v+1+\sqrt{(u-v-1)^2+4uv}}{2u}.
\end{equation}
If one takes an initial state consisting of $N-1$ consecutive shocks
(see Figure 1) with the densities $\rho_1$, $\rho_2$, $\cdots$, and
$\rho_{N}$, it turns out that the shock positions have simple random
walk dynamics provided that the
$\frac{\rho_{i+1}(1-\rho_{i})}{\rho_{i}(1-\rho_{i+1})}=x^{-1}$ for
$i=1,\cdots,N-1$ and that
\begin{eqnarray}
&& \rho_{1}=\frac{1}{1+\kappa_{+}(\alpha,\gamma)},\\
&& \rho_{N}=\frac{\kappa_{+}(\beta,\delta)}{1+\kappa_{+}(\beta,\delta)}.
\end{eqnarray}
It is interesting to note that by choosing an appropriate form for
the auxiliary operators $\bar{D_{0}}$ and $\bar{D_{1}}$ an
$N$-dimensional representation of the PASEP's quadratic algebra can
be written as the following form
$$\tiny{
D_{0}=\left( \begin{array}{ccccc}
(1-\rho_{N}) & 0& \cdots& 0 &0\\
1&\frac{\delta_{r,N-1}}{\delta_{l,N-1}}(1-\rho_{N-1})&
\cdots&0&0\\0&1& \cdots&0&0\\
\vdots &\vdots& \ddots&\vdots&\vdots\\
0&0& \cdots&\frac{\delta_{r,N-1}}{\delta_{l,N-1}}
\times\cdots\times\frac{\delta_{r,2}}{\delta_{l,2}}(1-\rho_{2})&0\\
0&0& \cdots&1&\frac{\delta_{r,N-1}}{\delta_{l,N-1}}
\times\cdots\times\frac{\delta_{r,1}}{\delta_{l,1}}(1-\rho_{1})
\end{array} \right)}\\
$$
$$\tiny{
D_{1}=\left( \begin{array}{ccccc}
\rho_{N} & 0& \cdots& 0 &0\\
0&\frac{\delta_{r,N-1}}{\delta_{l,N-1}}\rho_{N-1}& \cdots&0&0\\
0&0& \cdots&0&0\\
\vdots &\vdots& \ddots&\vdots&\vdots\\
0&0& \cdots&\frac{\delta_{r,N-1}}{\delta_{l,N-1}}
\times\cdots\times\frac{\delta_{r,2}}{\delta_{l,2}}\rho_{2}&0\\
0&0& \cdots&0&\frac{\delta_{r,N-1}}{\delta_{l,N-1}}\times\cdots
\times\frac{\delta_{r,1}}{\delta_{l,1}}\rho_{1}
\end{array} \right)}\\
$$
in a basis in which $D_{1}$ is diagonal but not the $D_{0}$. In this
matrix representation the quantities of $\delta_{r,i}$ and $\delta_{l,i}$
for $i=1,\cdots,N-1$ are the hopping rates of the $i$th shock position to
the right and to the left respectively. This can easily be done by using a
similarity transformation. It is worth mentioning that this representation
is exactly the one introduced in \cite{ms}; however, it is written here in
terms of the densities of the shocks $\rho_{1},\cdots,\rho_{N}$ and their
hopping rates $\delta_{r,i}$ and $\delta_{l,i}$ for $i=1,\cdots,N-1$. It can
be easily verified that by taking $N=2$ and  applying a similarity transformation
the two-dimensional representation (\ref{REP}) can be recovered. One should note that
the eigenvalues of the matrices are not changed under similarity transformations. \\
As another example, we consider the BCRW as a two-states system with
open boundaries and the following non-vanishing rates
\begin{equation}
w_{34},\quad w_{24},\quad w_{42},\quad w_{43},\quad w_{32},\quad
w_{23},\quad \alpha, \quad \gamma,\quad \beta
\end{equation}
and $\delta$ equal to zero. The quadratic algebra of the BCRW is
\begin{equation}
\label{BCRWalg}
\begin{array}{l}
\bar{D_{0}} D_{0}- D_{0}\bar{D_{0}}=0\\
\omega_{23}D_{1}D_{0}+\omega_{24}D_{1}^2-(\omega_{32}+\omega_{42})
D_{0}D_{1}=\bar{D_{0}}D_{1}-D_{0}\bar{D_{1}}\\
-(\omega_{23}+\omega_{43})D_{1}D_{0}+\omega_{34}D_{1}^2+\omega_{32}
D_{0}D_{1}=\bar{D_{1}}D_{0}-D_{1}\bar{D_{0}}\\
\omega_{43}D_{1}D_{0}-(\omega_{24}+\omega_{34})D_{1}^2+\omega_{42}
D_{0}D_{1}=\bar{D_{1}}D_{1}-D_{1}\bar{D_{1}}\\
\langle W \vert (\alpha D_{0}-\gamma D_{1})=\langle W
\vert\bar{D_{0}}=-\langle W \vert\bar{D_{1}}\\
\beta D_{1} \vert V \rangle =\bar {D_{0}} \vert V \rangle =-\bar
{D_{1}} \vert V \rangle.
\end{array}
\end{equation}
It is known that for the BCRW the dynamics of a single product shock
measure under the Hamiltonian of the system is a random walk
provided that \cite{kjs}
\begin{equation}
\label{BCRWcond}
\begin{array}{l}
\frac{1-\rho_1}{\rho_1}=\frac{\omega_{24}+\omega_{34}}{\omega_{42}+\omega_{43}}\\
\frac{1-\rho_1}{\rho_1}=\frac{\omega_{23}}{\omega_{43}}\\
\gamma= \frac{1-\rho_1}{\rho_1}\alpha + (1-\rho_1) \omega_{32}-
\frac{1-\rho_1}{\rho_1} \omega_{43} + \rho_1 \omega_{34}.
\end{array}
\end{equation}
In this case the density of the particles at the right hand side of
the shock position is zero $\rho_2=0$. One can easily check that the
following two-dimensional representation
\begin{equation}
D_{0}=\left( \begin{array}{cc}
1 & 0\\
d_{0}&\frac{\delta_{r}}{\delta_{l}}(1-\rho_{1})\\
\end{array} \right),\\
D_{1}=\left( \begin{array}{cc}
0 & 0\\
-d_0& \frac{\delta_{r}}{\delta_{l}}\rho_{1}\\
\end{array} \right)
\end{equation}
with $\delta_{r}=\frac{w_{43}}{\rho_{1}}$ and
$\delta_{l}=(1-\rho_{1})w_{32}+\rho_{1} w_{34}$ satisfies
(\ref{BCRWalg})
provided that the constraints (\ref{BCRWcond}) are satisfied.\\
For the AKGP, the non-vanishing parameters are $\omega_{12}$,
$\omega_{13}$, $\omega_{42}$, $\omega_{43}$, $\omega_{32}$, $\alpha$
and $\beta$. The particles are allowed to enter the system only from
the first site with the rate $\alpha$ and leave it from the last
site of the lattice with the rate $\beta$. In this case we have
$\rho_1=1$ and $\rho_2=0$. There are no additional constraints on
the rates for this model. The dynamics of shock measure generated by
the Hamiltonian of the system will be a simple random walk on the
lattice and the shock position hopping rates are $\delta_l=\omega_{13}$
and $\delta_r=\omega_{43}$ \cite{kjs}. The quadratic algebra of the AKGP is given by
\begin{equation}
\label{AKGPalg}
\begin{array}{l}
\omega_{13}D_{1}D_{0}+\omega_{12}D_{0}D_{1}= \bar{D_{0}} D_{0}-D_{0} \bar{D_{0}}\\
-(\omega_{12}+\omega_{32}+\omega_{42}) D_{0}D_{1}=\bar{D_{0}}D_{1}-D_{0}\bar{D_{1}}\\
-(\omega_{13}+\omega_{43})D_{1}D_{0}+\omega_{32}D_{0}D_{1}=\bar{D_{1}}D_{0}-D_{1}\bar{D_{0}}\\
\omega_{43}D_{1}D_{0}+\omega_{42}D_{0}D_{1}=\bar{D_{1}}D_{1}-D_{1}\bar{D_{1}}\\
\langle W \vert \alpha D_{0}=\langle W \vert\bar{D_{0}}=-\langle W \vert\bar{D_{1}}\\
\beta D_{1} \vert V \rangle =\bar {D_{0}} \vert V \rangle =-\bar
{D_{1}} \vert V \rangle
\end{array}
\end{equation}
which has a two-dimensional representation given by
\begin{equation}
D_{0}=\left( \begin{array}{cc}
1 & 0\\
d_{0}&0\\
\end{array} \right),\\
D_{1}=\left( \begin{array}{cc}
0 & 0\\
-d_0&\frac{\delta_{r}}{\delta_{l}}\\
\end{array} \right).
\end{equation}
One can also apply our procedure to the systems with $n>2$. In what
follows we investigate an open system with three-states at each
lattice site associated with two different types of particles
(denoted by $1$ and $2$) and vacancies (denoted by $0$). The
particles of different types are allowed to enter and leave the
lattice from the boundaries. There is also a probability for
changing the particle type at the boundaries. A couple of models of
this type have already been studied in \cite{jm,rs,ja1,ts1,ts2} and
shown that a traveling wave solution might develop in the system. In
this case the stochastic dynamics are defined in terms of 72
transition rates $w_{ij}$ ($i\neq j$) where $i,j=1,\cdots,9$. In the
basis ($00,01,02,10,11,12,20,21,22$) we will consider the case
studied in \cite{ts2} in which the nonzero transition rates are
\begin{equation}
w_{24},\quad w_{42},\quad w_{37},\quad w_{73},\quad w_{86},\quad w_{68},\quad w_{61},\quad w_{16},\quad w_{18},\quad w_{81}.
\end{equation}
For injection and extraction of particles at the left boundary we introduce
the rates :
\begin{eqnarray}
  1 \rightleftharpoons 0
   \quad & \alpha_1, \; \gamma_1, \nonumber \\
  1\rightleftharpoons 2
  \quad & \alpha_2,\; \gamma_2,  \nonumber \\
 0 \rightleftharpoons 2
   \quad & \alpha_3, \; \gamma_3,
\end{eqnarray}
and for the right boundary
\begin{eqnarray}
  1 \rightleftharpoons 0
   \quad & \delta_1, \; \beta_1, \nonumber \\
  1\rightleftharpoons 2
  \quad & \delta_2,\; \beta_2,  \nonumber \\
 0 \rightleftharpoons 2
   \quad & \delta_3, \; \beta_3.
\end{eqnarray}
The quadratic algebra of the system in this case is given by
\begin{equation}
\label{THREEalg}
\begin{array}{l}
-(w_{61}+w_{81})D_{0}D_{0}+w_{16}D_{1}D_{2}+w_{18}D_{2}D_{1}=\bar{D_{0}}D_{0}-D_{0}\bar{D_{0}}\\
-w_{42}D_{0}D_{1}+w_{24}D_{1}D_{0}=\bar{D_{0}}D_{1}-D_{0}\bar{D_{1}}\\
-w_{73}D_{0}D_{2}+w_{37}D_{2}D_{0}=\bar{D_{0}}D_{2}-D_{0}\bar{D_{2}}\\
w_{42}D_{0}D_{1}-w_{24}D_{1}D_{0}=\bar{D_{1}}D_{0}-D_{1}\bar{D_{0}}\\
w_{61}D_{0}D_{0}-(w_{16}+w_{86})D_{1}D_{2}+w_{68}D_{2}D_{1}=\bar{D_{1}}D_{2}-D_{1}\bar{D_{2}}\\
w_{73}D_{0}D_{2}-w_{37}D_{2}D_{0}=\bar{D_{2}}D_{0}-D_{2}\bar{D_{0}}\\
w_{81}D_{0}D_{0}+w_{86}D_{1}D_{2}-(w_{18}+w_{68})D_{2}D_{1}=\bar{D_{2}}D_{1}-D_{2}\bar{D_{1}}\\
\bar{D_{2}}D_{2}-D_{2}\bar{D_{2}}=\bar{D_{1}}D_{1}-D_{1}\bar{D_{1}}=0\\
\langle W \vert (\gamma_{1}D_{0}-(\alpha_{1}+\alpha_{2})D_{1}+\gamma_{2}D_{2}+\bar{D_{1}})=0\\
\langle W \vert (\alpha_{3}D_{0}+\alpha_{2}D_{1}-(\gamma_{2}+\gamma_{3})D_{2}+\bar{D_{2}})=0\\
(\beta_{1}D_{0}-(\delta_{1}+\delta_{2})D_{1}+\beta_{2}D_{2}-\bar{D_{1}})\vert V \rangle =0\\
(\delta_{3}D_{0}+\delta_{2}D_{1}-(\beta_{2}+\beta_{3})D_{2}-\bar{D_{2}})\vert V \rangle =0
\end{array}
\end{equation}
It has been shown that the shocks of the form (\ref{PSM}) can
develop in the system provided that some constraints are fulfilled.
The shock in this model only hops to the left i.e. $\delta_{l}\neq
0$ and $\delta_{r}=0$. In the following we show that the steady
state of the system can be written as a matrix product state using
two-dimensional representations of the quadratic algebra
(\ref{THREEalg}). We will study two different cases. In the first
case we assume $w_{42}=w_{18}=w_{68}=0$ and also $\rho_{2}^{(1)}=1,
\rho_{2}^{(2)}=0.$ The two densities $\rho_{1}^{(1)}$ and
$\rho_{1}^{(2)}$ are free parameters. The matrix representation of
the algebra (\ref{THREEalg}) is now given by
\begin{equation}
\label{REP1}
D_{0}=\left( \begin{array}{cc}
0 & 0\\
d_{0}&0\\
\end{array} \right),\\
D_{1}=\left( \begin{array}{cc}
1 & 0\\
d_{1}&0\\
\end{array} \right),\\
D_{2}=\left( \begin{array}{cc}
0 & 0\\
d_{2}&0\\
\end{array} \right)
\end{equation}
in which $d_{0}+d_{1}+d_{2}=0$. As can be seen the two operators
$D_{0}$ and $D_{2}$ commute with each other but not necessarily with
$D_{1}$. The auxiliary operators $\bar{D_{0}}$, $\bar{D_{1}}$ and
$\bar{D_{2}}$ are taken to be zero. The matrix element $w_{2}$ in
(\ref{REP}) is zero in this case and the following constraints have
to be fulfilled
\begin{equation}
\delta_{1}=\delta_{2}=\alpha_{1}=\alpha_{2}=0.
\end{equation}
The rest of the boundary rates besides $d_{1}$ and $d_{2}$ should satisfy two linear equations
\begin{eqnarray}
&&\beta_{1} d_{1}+(\beta_{1}-\beta_{2})d_{2}=0\\
&&(\beta_{1}+\delta_{3})d_{1}+(\beta_{1}+\beta_{3}+\delta_{3})d_{2}=0.
\end{eqnarray}
Assuming $d_{1},d_{2} \neq 0$ then those boundary rates which satisfy
$\beta_{1}\beta_{3}+\beta_{1}\beta_{2}+\beta_{2}\delta_{3}=0$ generate a solution.\\
In the second case we consider $\rho_{1}^{(2)}=\rho_{2}^{(2)}=0$, $\rho_{2}^{(1)}=1$
and $\rho_{1}^{(1)}$ is a free parameter. The steady state of the system in this case is given by
following two-dimensional matrices
\begin{equation}
\label{REP2}
D_{0}=\left( \begin{array}{cc}
0 & 0\\
d_{0}&0\\
\end{array} \right),\\
D_{1}=\left( \begin{array}{cc}
1 & 0\\
-d_{0}&0\\
\end{array} \right),\\
D_{2}=\left( \begin{array}{cc}
0 & 0\\
0&0\\
\end{array} \right)
\end{equation}
which is similar to (\ref{REP1}) except that $d_{2}=0$. This means
that the probability of finding second-class particles is zero in
the steady state. Taking all the auxiliary operators $\bar{D_{0}}$,
$\bar{D_{1}}$ and $\bar{D_{2}}$ equal to zero, one can simply check
that (\ref{REP2}) satisfy the algebra (\ref{THREEalg}) provided that
$w_{42}=0$ and
\begin{equation}
\delta_{1}=\delta_{2}=\alpha_{1}=\alpha_{2}=\beta_{1}=\delta_{3}=0.
\end{equation}
In this paper we have introduced a general procedure that can be
used to investigate those one-dimensional multi-species system with
open boundaries whose steady states can be written as a
superposition of single product shock measures of the form
(\ref{PSM}) with random walk dynamics. Instead of applying the
Hamiltonian of such systems on a single product shock measure
(\ref{PSM}) as an initial configuration and looking for the
conditions under which this initial configuration evolves into a
linear combination of product shock measures with different shock
positions, as it is done in \cite{rs,ts1,ts2,ps}, one can simply
follow the following steps:
\begin{enumerate}
\item{Using the standard MPF we find the quadratic algebra associated
with the one-dimensional open boundaries system.}
\item{We investigate whether or not it has two-dimensional matrix
representations and that if these representations have the same
structure as (\ref{REP}).}
\item{If the answer to these questions is positive, we then calculate
the coefficients $c_{k}$'s in (\ref{SSS}) using (\ref{COEF4}).}
\end{enumerate}
We should emphasis that one cannot use any arbitrary matrix
representation to generate the coefficients $c_{k}$'s using
(\ref{COEF4}). Only when the two-dimensional matrix representation
of the quadratic algebra is of the form (\ref{REP}) the coefficients
$c_{k}$'s have the right structure (\ref{COEF1})-(\ref{COEF3}). \\
It would be interesting to investigate if this procedure also works
for the systems with long range interactions \cite{ps} or even for
the systems in which multiple product shock measures might evolve in
them. As we have shown the generic form of the matrix representation
(\ref{REP}) still holds for the PASEP with multiple shocks; however,
it is not clear if it is true for other systems in general. It is
also interesting to investigate if this property holds for the
models defined a ring geometry or in the presence of second class
particles. This is under our investigations. \\
F. H. J. would like to thank  V. Rittenberg, G. M. Sch\"utz and R.
B. Stinchcombe for useful discussions and comments.

\end{document}